\documentstyle[psfig,amssymb]{mn}

\title[Star formation in spiral galaxies]
{A test of arm--induced star formation in spiral galaxies from
near--IR and H$\alpha$ imaging}

\author[M.S. Seigar \& P.A. James] 
{Marc S. Seigar$^1$ and Phil A. James$^2$\\  
$^1$Joint Astronomy Centre, 660 N. A'ohoku Place, Hilo, HI 96720, USA\\
{\tt email: m.seigar@jach.hawaii.edu}\\
$^2$Astrophysics Research Institute, Liverpool 
John Moores University, Egerton Wharf, Birkenhead, CH41 1LD, UK\\
{\tt email: paj@astro.livjm.ac.uk}\\
}

\begin{document} 
\maketitle

\begin{abstract}
We have imaged a sample of 20 spiral galaxies in H$\alpha$
and in the near--infrared {\em K} band (2.2$\mu$m), in order to
determine the location and strength of star formation in these objects
with respect to perturbations in the old stellar population. We have
found that star formation rates are significantly enhanced in the
vicinity of {\em K} band arms. We have also found that this 
enhancement in star formation rate in arm regions correlates well
with a quantity that measures the relative strengths of shocks in
arms. Assuming that the {\em K} band light is
dominated by emission from the old stellar population, this shows that
density waves trigger star formation in the vicinity of spiral arms.

\end{abstract}
  

\section{INTRODUCTION}

Two of the leading theories of star formation in spiral galaxies use
the concept of a density wave, either as the actual triggering
mechanism, or as a means of organisation of star--forming material. The
first of these theories has been termed the large scale galactic shock
scenario (Roberts 1969; see also Shu et al. 1972; Tosa 1973; Woodward
1975; Nelson \& Matsuda 1977). This model hypothesises that the gas
settles into a quasi--stationary state, with a velocity and density
distribution that is driven by the gravitational field of the galaxy. The gas
response can be non--linear to an imposed azimuthal sinusoidal
potential, if relative motion between the density wave and the cold
interstellar medium (ISM) is supersonic (Binney \& Tremaine 1987). 
This leads to the formation
of a shock near the trailing edge of spiral arms, assuming that the
region is inside the corotation radius, which compresses the gas to
densities at which stars can form. In observations of spiral galaxies
the shock is thought to be characterised by dust--lanes seen on the
trailing edges of arms. The time delay needed for the onset of star
formation after the compression of the gas implies that star--forming
regions should be seen towards the leading edges of arms.

Many of the recent advances in this area have arisen from studies of
the atomic and molecular gas in nearby spiral galaxies,
through HI and CO line emission (e.g Nakai et al. 1994; Rand
1993, 1995). These studies reveal streaming velocities of gas through the 
spiral arms and find offsets between the
peaks of the gas density and the old stellar population, in 
agreement with the predictions of the large scale shock scenario.

The alternative picture to this form of triggered star formation is
stochastic star formation. \"Opik (1953) first hypothesised that a
supernova explosion could trigger star formation. A model proposed and
developed by several authors (e.g. Gerola \& Seiden 1978;
Seiden \& Gerola 1982; Seiden 1983; Jungwiert \& Palous 1994; Sleath
\& Alexander 1995, 1996) suggests that the dominant process for
forming stars is stochastic self--propagating star formation, and not
density wave triggering. In this model, density waves are only
responsible for the organisation of the ISM and stars, and for
concentrating new HII regions along spiral arms (Elmegreen \&
Elmegreen 1986; Elmegreen 1993). Thus, star formation rate {\em
efficiency} (i.e. normalised to unit mass of disc material) should be
unaffected by location in arm or interarm regions, in this model.  This
is in clear distinction to the predictions of the large--scale shock
model.

Previous H$\alpha$ studies of spiral galaxies (Kennicutt 1989, 1998a;
Kennicutt, Tamblyn \& Congdon 1994; and the review by Kennicutt 1998b)
have mainly looked at global star--formation, particularly the form of
the Schmidt law (Schmidt 1959, 1963). In this paper we are looking at
localised star formation in disc galaxies as well as global properties
of star formation, and comparing the distribution of star formation
with the underlying old stellar population. The main aim of this paper
is the analysis of star formation efficiencies in arm and interarm
regions in spiral galaxies.

Other studies of star formation efficiencies include Lord \& Young (1990) 
and Tacconi \& Young (1986). Lord \& Young (1990) looked at the molecular, 
neutral, and ionized hydrogen distributions in M51. They compared a ratio
of massive star formation rates (MSFR) to gas surface density between arm
regions and interarm regions and found a higher ratio in the arm regions. 
Tacconi \& Young (1986) performed a similar analysis for NGC 6946 and also
found that star formation is more efficient in arm regions than in interarm
regions.
Cepa \& Beckman (1990) and Knapen et al. (1992) again looked at star formation
efficiencies in arm and interarm regions, but with high spatial resolution.
Cepa \& Beckman (1990) compare the H$\alpha$/$\sigma_{HI}$ ratio (where
$\sigma_{HI}$ is the HI surface brightness) in the 
arm and interarm
regions of NGC 3992 and NGC 628. Knapen et al. (1992) compare the 
H$\alpha$/$\sigma(H_2+HI)$ ratio in arm and interarm regions. Both Knapen et
al. (1992) and Cepa \& Beckman (1990) explicitly find evidence for spiral
arm triggering of star formation, via a non--linear dependence of SFR on
gas density.
Wyder, Dolphin \& Hodge (1998) obtained HST WFPC2  V band (F555W)
and I band (F702W) data of the spiral galaxy NGC 4321. From the I--V colour
of the arm and interarm regions, they concluded that the SFR over the last 5 
Myr has been approximately 4 times larger in the arm regions than in the
surrounding interarm regions.
Also, Knapen (1998) performed a study of HII regions in M100, using a new
H$\alpha$ image. He found that the arms collect HII regions in a similar way
to that described by Elmegreen \& Elmegreen (1986). He concluded that spiral
arms may trigger some star formation, but do not affect the HII region
luminosity function or mass distribution. Finally, Alonso--Herrero \& Knapen
(2001) used the specific SFR or SFR per stellar mass (i.e. a comparison
of Pa$\alpha$ flux with {\em H} band continuum emission), and found that while
the specific SFR in the central region does not vary statistically with galaxy
type, it is higher in barred than in non-barred galaxies.

The remainder of this paper is arranged as follows. Section 2
describes the observations and data analysis; 
section 3 describes local
star--formation rates with respect to properties of spiral structure;
and section 4 contains our conclusions.

\section{OBSERVATIONS AND DATA REDUCTION}

In this sample, 14 of the galaxies were selected as part of a program 
designed to understand the underlying structure of galaxies. The selection 
of these galaxies, and near--IR observations and data reduction, is
described in Seigar \& James (1998a).  Briefly, the full galaxy sample
is comprised of 45 spirals of types Sa--Sdm, with inclinations $\lse$
45$^{\circ}$ and diameters $\lse$1\farcm5, the latter enabling them to
be imaged in the field of the UKIRT near--IR camera IRCAM3. 
A further 5 galaxies have been observed with the UKIRT Fast Track Imager
(UFTI), with the same selection criteria. The final galaxy was observed
with INGRID on the William Herschel Telescope (WHT), again with the same
selection criteria, except its diameter was chosen to be $\le$4$^\prime$.

The optical data
presented here are broad--band (R) and narrow--band
(redshifted H$\alpha$) images of the 20 galaxies described above.
The observations were made using the 1.0m Jacobus Kapteyn Telescope
(JKT) on La Palma. Some of the galaxies were observed 
on the nights 1998 May 14--20, with some additional
observations being taken during a later run, 1999 February 4--11,
using the 1024$\times$1024 Tek4 CCD camera which has
0\farcs331 pixels and a field of 5\farcm65$\times$5\farcm65.
The rest of the objects were observed
between February 2000 and January 2002, using the 2048$\times$2048 SITe2
CCD camera, which has 0\farcs33 pixels and a field of 
11\farcm3$\times$11\farcm3. The
filters used were standard Harris R filter, and narrow--band filters
with the following central wavelengths in Angstroms (full--width at
half maximum in brackets): 6594(44), 6626(44), 6656(44), 6686(44),
6695(53), 6712(42), 6727(48), 6785(58).  Measured transmission curves
are available for all these filters, and were used in selecting the
optimum filter for each galaxy redshift, and in flux calibrating the
images, as described below.

All 20 galaxies were observed in redshifted H$\alpha$ filters, with
between one and four 1200 second integrations for each galaxy.  All were
also observed at R with integration times of 300--900 seconds. In
addition, the following standard stars, selected from Landolt (1992),
were observed: PG1047+003 A, B \& C; PG1323--086 A, B \& C; PG1633+099
B, C \& D; SA111--773; and SA111--775.  Two spectrophotometric
standards from Oke (1990), Feige~34 and G191B2B, were also observed.

Data reduction made use of standard STARLINK image reduction packages.
All images were bias subtracted, and then flat fielded using twilight
sky flats.  There was no evidence for any fringing effects in broad--
or narrow--band images.  The H$\alpha$ images were continuum--subtracted
using scaled {\em R} band images.  The scaling factor applied to the
{\em R} band images was calculated photometrically, using the ratio of
fluxes of red standard stars from the Landolt (1992) list.  These
values were checked using the spectrophotometric standards from Oke
(1990), to ensure that the red standards did not have strong spectral
features in the passband of the H$\alpha$ filter, and by numerically
integrating under the filter profiles.  All methods gave very similar
results, but the red standards were adopted as best representing the
colour of the galaxy continuum light. The throughputs of the narrow
band filters to continuum light were measured in this way to be
between 36 and 64 times lower than that of the {\em R} band filter.
The narrow--band images were continuum subtracted by normalising broad--
and narrow--band images to the same integration time, dividing the
former by the ratio of the throughputs, and subtracting this scaled
broad--band image.

Calibration of the H$\alpha$ photometry required calculation of the
effective throughput of the narrow--band filter to H$\alpha$ light.
This was done by calculating the throughputs of both narrow-- and
broad--band filters at the wavelength corresponding to redshifted
H$\alpha$, for each galaxy, from the measured transmission curves. The
narrow--band transmission was then corrected by subtracting the
broad--band transmission, scaled by the ratio of continuum throughputs
(36--64, as given above).  Given the large size of this factor, the
continuum subtraction works well,  {\em even though the R band filter
contains the H$\alpha$ line}, and the only effect of this is to reduce
the effective throughput for H$\alpha$ emission by a few percent.
Effective throughputs calculated in this way were in the range
28--50\%, typically $\sim$45\%.

This continuum subtraction procedure worked extremely well, with the
only problems being flat--fielding residuals which were apparent in
some of the continuum ({\em R} band) images taken in bright moonlight, and
a slight change in plate--scale between the broad-- and narrow--band
images, which was not significant over the size of the galaxies in
this study.  The accuracy of the continuum subtraction was confirmed
by the almost complete removal of old--stellar light (non--star--forming
bulges etc) in the resulting line images.  Foreground stars were also removed
fairly accurately, but residuals of a few percent were sometimes left
due to the differing colours compared with integrated galaxy light.
There was also a small residual sky level, which was subtracted before
photometric measurements were made.

The photometry was obtained using the STARLINK software package GAIA.
This extracted H$\alpha$ fluxes as raw counts, which were then
corrected for airmass, using a correction based on the {\em R} band
observations of the standard stars listed above.  A correction was
then applied for the effective throughput of the narrow--band filter to
the redshifted H$\alpha$ radiation as described above.

\begin{table}
\caption{Morphological parameters for the galaxy sample.}
\begin{center}
\begin{tabular}{l l l l }
\hline
Galaxy		& Galaxy	& V$_{rec}$     & Telescope/instrument 	\\  
name		& Classn.	& (km~s$^{-1}$) &			\\
\hline	
IC 742    & SBab  & 6425  & UKIRT/IRCAM3	\\
NGC~2628  & SABc  & 3622  & UKIRT/IRCAM3	\\
NGC~5737  & SBb   & 9517  & UKIRT/IRCAM3	\\
NGC~6347  & SBb   & 6144  & UKIRT/IRCAM3	\\
NGC~6379  & Scd   & 5973  & UKIRT/IRCAM3	\\
NGC~6574  & SABbc & 2282  & UKIRT/IRCAM3	\\
UGC~3053  & Scd   & 2407  & UKIRT/IRCAM3	\\
UGC~3171  & SBcd  & 4553  & UKIRT/IRCAM3	\\
UGC~3296  & Sab   & 4266  & UKIRT/IRCAM3	\\
UGC~3578  & SBab  & 4531  & UKIRT/IRCAM3	\\
UGC~3936  & SBbc  & 4725  & UKIRT/IRCAM3	\\
UGC 4270  & SABbc & 2479  & UKIRT/UFTI		\\
UGC 4705  & SBb   & 2526  & UKIRT/UFTI		\\
UGC 4779  & SAc   & 1289  & WHT/INGRID		\\
UGC~5434  & SABb  & 5580  & UKIRT/IRCAM3	\\
UGC 5786  & SABbc & 993   & UKIRT/UFTI		\\
UGC 6132  & SBb   & 979   & UKIRT/UFTI		\\
UGC~6332  & SBa   & 6245  & UKIRT/IRCAM3	\\
UGC 7985  & SABd  & 652   & UKIRT/UFTI		\\
UGC~11524 & Sc    & 5257  & UKIRT/IRCAM3	\\
\hline
\end{tabular}
\end{center}
\end{table}

Table 1 contains the following
entries: Galaxy name (column 1), galaxy classification (column 2),
heliocentric redshift in km~s$^{-1}$ (column 3) and the telescope
and instrument with which they were observed (column 4).

\section{RESULTS}

In this section we look at how density waves affect the local star
formation rates {\em within} the discs of individual galaxies,
even though other factors such as cold gas mass or galaxy environment
may dominate the differences in star formation rate {\em between}
galaxies.  To test the local effect of arms, we compared the degree of
concentration of H$\alpha$ and {\em K} band light in the arms of the present
galaxy sample. We use the H$\alpha$ data as an indicator of star formation
rate and the 
{\em K} band data as a tracer of where the density waves are (the {\em K}
band largely reflects old stars, which is a good tracer of the overall stellar
mass density, and hence of density waves within discs).
If the simple model of Elmegreen (1993) is true, the
density waves should simply gather up all types of disc material to
the same extent, and the same arm--to--interarm contrast should be seen
in all disc tracers.  Then, arms would be just as prominent in the
near--IR emission from old stars as in the H$\alpha$ emission from
star--formation regions.  If, however, spiral density waves trigger
star formation to a significant extent, then the arm--to--interarm
contrast could be substantially enhanced in the star--formation tracer
relative to the old stellar population.  Note that this rather naive
argument neglects the effects of extinction, but since the optical
light is likely to be more affected than the near--IR, any relative
enhancement in H$\alpha$ arm contrasts compared to say the {\em K} band
should be a strong indication of arm--induced star formation. 

\subsection{A comparison of H$\alpha$/K ratios for arm and interarm regions}

In order to measure any enhancement in H$\alpha$ arm contrasts compared to 
{\em K} band arm contrasts,
we came up with the following conservative test, which is based
on the technique used to measure star--formation efficiency by Knapen et al.
(1996), to identify arm--induced star formation in the present sample
of galaxies.  For those galaxies which showed well--defined spiral arms
in their {\em K} band images, we rebinned the {\em K} images to the same
pixel scale as the H$\alpha$ images, and shifted them to overlay the
H$\alpha$ images, using stellar centroids as references.  This
procedure should be accurate to a fraction of a pixel, or better than
0\farcs1. We then used the `ARD region' option within the STARLINK
GAIA package to define a polygon lying around each {\em K} band arm,
without reference to any of the optical images.  Elliptical regions
were also fitted by eye to define the disc of the galaxy, excluding
the central bulge, and extending to the outer ends of the spiral arms.
The images were sky subtracted, and then the number of detected counts
within each of the arm polygons, and the overall disc, were
calculated.  GAIA then permits the apertures to be imported into the
H$\alpha$ images, which had been aligned exactly, such that the
apertures lie over the same physical regions of the galaxies at both K
and H$\alpha$.  Photometry was then obtained for all regions (arms,
total disc and nucleus) in the H$\alpha$ images in exactly the same
way as for the {\em K} images. Finally, ratios were calculated of H$\alpha$
flux divided by {\em K} band flux for all of the regions. The test is
then to see whether this ratio is larger for the arm regions than for
the disc generally, where the latter ratio is taken from the total
disc minus the central region.  Any obvious foreground stars are
removed from from the arm and disc regions before this comparison is
done.

\begin{table*}
\caption{Normalised H$\alpha$/{\em K} ratios for 49 arm regions in 20 galaxies.}
\begin{center}
\begin{tabular}{l l l l l l}
\hline
Galaxy name	  &  Arm 1 & Arm 2 & Arm 3 & Arm 4 & Average  \\  
\hline	
IC~742    & 1.19$\pm$0.12   & 1.48$\pm$0.13   & -- & -- & 1.33$\pm$0.18 \\
NGC~2628  & 0.95$\pm$0.04   & 1.48$\pm$0.06   & 1.73$\pm$0.10   & 1.32$\pm$0.08 & 1.37$\pm$0.10\\
NGC~5737  & 1.45$\pm$0.07   & 1.12$\pm$0.12   & -- & -- & 1.29$\pm$0.12 \\
NGC~6347  & 1.89$\pm$0.08   & 1.19$\pm$0.06   & -- & -- & 1.54$\pm$0.10 \\
NGC~6379  & 3.35$\pm$0.18   & 1.81$\pm$0.13   & 0.77$\pm$0.13   & -- & 1.98$\pm$0.13 \\
NGC~6574  & 1.17$\pm$0.09   & 1.39$\pm$0.09   & 0.87$\pm$0.09   & 1.67$\pm$0.09 & 1.28$\pm$0.09 \\
UGC~3053  & 0.78$\pm$0.06   & 0.93$\pm$0.09   & 0.58$\pm$0.06   & -- & 0.76$\pm$0.06 \\
UGC~3171  & 1.21$\pm$0.15   & 2.59$\pm$0.21   & 1.21$\pm$0.08   & 1.56$\pm$0.12  & 1.64$\pm$0.12 \\
UGC~3296  & 1.03$\pm$0.12   & 1.68$\pm$0.08   & -- & -- & 1.36$\pm$0.12 \\
UGC~3578  & 1.95$\pm$0.22   & 2.14$\pm$0.16   & -- & -- & 2.05$\pm$0.19 \\
UGC~3936  & 1.23$\pm$0.11   & 1.01$\pm$0.09   & 0.90$\pm$0.08   & -- & 1.05$\pm$0.09 \\
UGC 4270  & 1.02$\pm$0.10   & 1.20$\pm$0.05   & -- & -- & 1.10$\pm$0.08	\\
UGC 4705  & 2.49$\pm$0.41   & 0.94$\pm$0.19   & -- & -- & 1.72$\pm$0.32 \\ 
UGC 4779  & 1.71$\pm$0.19   & 2.29$\pm$0.27   & -- & -- & 2.00$\pm$0.23 \\
UGC~5434  & 1.79$\pm$0.13   & 0.86$\pm$0.05   & -- & -- & 1.32$\pm$0.12 \\
UGC 5786  & 2.17$\pm$0.29   & 9.87$\pm$1.86   & -- & -- & 6.02$\pm$0.54 \\
UGC 6132  & 1.53$\pm$0.23   & 2.20$\pm$0.32   & -- & -- & 1.87$\pm$0.25 \\
UGC~6332  & 2.95$\pm$0.09   & 0.32$\pm$0.08   & -- & -- & 1.63$\pm$0.29 \\
UGC 7985  & 2.34$\pm$0.27   & 1.90$\pm$0.20   & -- & -- & 2.12$\pm$0.17 \\
UGC~11524 & 0.87$\pm$0.30   & 1.91$\pm$0.46   & -- & -- & 1.39$\pm$0.30 \\
\hline
\end{tabular}
\end{center}
\end{table*}

If arms represent regions where all disc material is concentrated by
an equal factor, then the H$\alpha$/{\em K} ratio would be the same in arms
as in the disc overall.  If there were no connection between star
formation and {\em K} band structure, then this ratio would be lower in the
selected arm regions, since these were chosen to have higher than
average {\em K} surface brightnesses.  However, in the majority of cases, we
found H$\alpha$/{\em K} ratios to be 
significantly higher in the {\em K} band arms
than in the discs generally.  We measured a total of 49 arm regions in
20 galaxies (see Table 2); 38 (76\%) of these arms had H$\alpha$/K
ratios greater than that of the disc of the same galaxy.  Averaging
the ratios over all arms observed in a given galaxy, we find (Table 2
final column) that 19 of the 20 galaxies show a net enhancement in
H$\alpha$/{\em K} ratio within their arm regions. The errors in table 2
were calculated by both adding and subtracting the uncertainty in the 
sky background from the H$\alpha$ images and performing the same analysis 
for the original H$\alpha$ image, the H$\alpha$ image subtracted by the
sky background uncertainty, and the H$\alpha$ image added by the sky 
background uncertainty. The error in the sky background was calculated using
both the formal uncertainty from pixel--pixel noise, and larger--scale
systematic variations in the sky structure. Combining the results for
all 49 arms, the median ratio was 1.47$\pm$0.11 times higher in the arms
than in the corresponding discs. We have taken the median ratio in order to
give less weight to galaxies which may be affected by starbursts. 
Given the conservative nature of the
test, this represents a highly significant finding of triggering of
star formation within spiral arms. We checked that the location of the
arms would not have been changed had we used {\em J} rather than {\em
K} images, and thus our results are not affected by the possibility
that hot dust may contribute to galaxy light at {\em K} (James \& Seigar
1999).

Two of the galaxies warrant further comment.  UGC~3053 exhibits 3
well--defined arm segments in its {\em K} band image, but there is
little or no H$\alpha$ luminosity from HII regions associated with
these arms. This is reflected in the low H$\alpha$/{\em K} ratios for all
three arm regions in this galaxy, which thus appears to have no
arm--induced star formation according to this test.  UGC~6332 has two
tightly wound arms, which could be classified as a single ring, and
which appear highly symmetrical in the {\em K} image.  However, this
symmetry breaks down completely in H$\alpha$ light, since one arm is
clearly associated with a string of HII regions, and the other with no
detectable HII regions at all. As a result, these two
arms give rise respectively to the second--highest, and to easily the
lowest, of the 49 measured H$\alpha$/{\em K} ratios. This asymmetry 
could be an extinction effect, due to the offset between the dust lane
and the peak of the induced star formation.

In this analysis we have ignored the result that the {\em K} band light may
have a contribution of up to 30\% from star formation (James \& Seigar
1999). An effect such as this will tend to increase the amplitude of the
{\em K} band spiral arms, and hence decrease the effect that we find. Hence our
result is conservative and can only be strengthened by making corrections
for this.

\subsection{The effect of arm strength on star formation rate}

\begin{figure}
\includegraphics{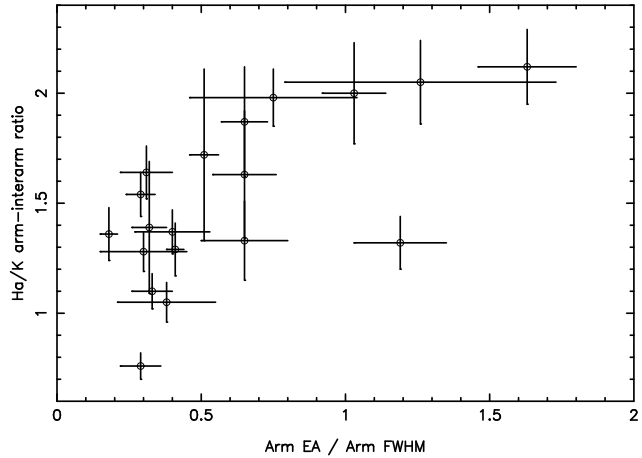}
\vspace*{6.5cm}
\caption{H$\alpha$/K ratio versus Arm EA/Arm FWHM ratio.}
\end{figure}

A further test of arm induced star formation is to see if the enhancement
in H$\alpha$ flux in the arm regions relates to the strength of shocks in
the arms. Seigar \& James (1998b) used the ratio between arm equivalent 
angle (EA) and arm full width half maximum (FWHM) as measured from {\em
K} band images, as a measure of the relative strength of shocks in spiral
arms. From here on, we refer to this ratio as the `central arm contrast'. 
Arm EA is defined as that angle subtended by the disc that contains
an amount of light equivalent to that in the spiral arm
(see Seigar \& James 1998a for a detailed discussion) and is therefore
a measure of the amount of light contained in a spiral arm. However, it 
contains no information about how concentrated the light from the arm is, and
this is the justification in dividing this quantity by arm FWHM (measured from
the cross--sectional light profiles of {\em K} band arms) to estimate
the `central arm concentration'. If
shocks are necessary to drive star formation, it seems reasonable that an arm
with a high central arm concentration will be more efficient at forming stars
than an arm with a low central arm concentration.
We have investigated the relationship between
this quantity and the H$\alpha$/K ratios measured in Table 2. 
The result is shown
in Figure 1. One galaxy is left out of this plot, UGC 5786, which is an
HII galaxy, and may
have processes other than arm induced star formation, contributing to its
extraordinarily high star formation rate.
Figure 1 shows a non--linear relationship between these two quantities, which 
is significant at the
99.93\% confidence level. It therefore seems that the star formation in spiral
galaxies is triggered by shocks in arms, up to a certain limiting threshold,
where H$\alpha$/K$\sim$2. 

\subsection{The star formation rate in UGC 5786}

Finally, UGC 5786 (NGC 3310) has a strong ring of star
formation at an average radius of 0.42$\pm$0.06 kpc from its centre.
The ring of star formation observed in this galaxy has been the
subject of many previous studies (e.g. Elmegreen et al. 2002).
Rings of nuclear star formation are typically associated with the
Inner Lindblad Resonance (ILR) and have been found in other galaxies,
e.g. M 74 (Wakker \& Adler 1995; James \& Seigar 1999). It should 
also be noted that UGC 5786 is classified as an HII galaxy. Bell
\& Kennicutt (2001) report an H$\alpha$ luminosity, $L_{H\alpha}=
3\times10^{41}$ ergs s$^{-1}$. Using the following equation,
\begin{equation}
SFR (M_{\odot} yr^{-1}) = 7.9\times10^{-42} L_{H\alpha} (ergs s^{-1})
\end{equation}
they calculated a SFR of 2.37 M$_{\odot}/$yr.
We have measured an $L_{H{\alpha}}=$(1.336$\pm$0.084)$\times$10$^{42}$ 
ergs s$^{-1}$. Also using equation
1, we have calculated a SFR of 10.55$\pm$0.66 M$_{\odot}/$yr. This is clearly
in disagreement with the SFR reported in Bell \& Kennicutt (2001). It should
be noted that we calculated a distance of 18.2 Mpc from a Virgocentric inflow
model (and using H$_{0}$=75 kms$^{-1}$Mpc$^{-1}$) for UGC 5786, 
whereas Bell \& Kennicutt (2001) use the distance calculated by Waller et al.
(1997), which is 13.9 Mpc. However, even taking into account this 
difference in measured distances to UGC 5786, 
our SFR is still a factor of $\sim$2.6 larger than that reported by
Bell \& Kennicutt (2001).

\section{CONCLUSIONS}

We have found an increase in H$\alpha$ flux in the vicinity of {\em K} band 
arms in this sample.
If the {\em K} band arms are dominated by light from the old stellar
population, then this can be interpreted as star formation triggered by a 
density wave. 
This result is in agreement with Lord \& Young (1990), who
found a higher star formation efficiency in arm regions. It also agrees
with Cepa \& Beckman (1990) and Knapen et al. (1992) who found evidence for
spiral arm triggering of star formation via a non--linear dependence of SFR on
gas density. 
This result therefore agrees with the large scale shock scenario (Roberts 1969;
Shu et al. 1972; Tosa 1973; Woodward 1975; Nelson \& Matsuda 1977), which 
predicts a higher star formation efficiency in arm regions. It does not
support the ideas of stochastic self--propagation star formation (e.g. Gerola
\& Seiden 1978; Seiden \& Gerola 1982; Seiden 1983; Jungwiert \& Palous 1994;
Sleath \& Alexander 1995, 1996) which predict no change in the star formation
efficiency from interarm to arm regions.

\section{ACKNOWLEDGMENTS}

The United Kingdom Infrared Telescope is operated by the Joint
Astronomy Centre on behalf of the U.K. Particle Physics and Astronomy
Research Council. The Jacobus Kapteyn Telescope and the William Herschel
Telescope are operated on the island 
of La Palma by the Isaac Newton Group in the Spanish Observatorio del 
Roque de los Muchachos of the Instituto de Astrofisica de Canarias.
This research has made use of the NASA/IPAC Extragalactic Database 
(NED) which is operated by the Jet Propulsion Laboratory, California 
Institute of Technology, under contract with the National Aeronautics 
and Space Administration. The authors thank Neville Shane 
for both helping out with the observations
and calculations provided in this paper. We also thank the referee for
comments which improved the content of this paper.

\end{document}